\documentclass[conference,A4,10pt]{IEEEtran}
\usepackage{blindtext, graphicx}
\usepackage{cite}

      \usepackage{graphicx}
\usepackage[cmex10]{amsmath}
\usepackage{amssymb}
\usepackage{amsthm}

\usepackage{algorithmic}
\usepackage[tight,footnotesize]{subfigure}
\usepackage{url}
\usepackage{algorithm}

\newcommand{\Hyp}{\mathsf{H}}
\newcommand{\Pro}{\mathsf{P}}
\newcommand{\Exp}{\mathsf{E}}
\newcommand{\cX}{\mathcal{X}}
\newcommand{\bcX}{\bar{\mathcal{X}}}

\newcommand{\cL}{\mathcal{L}}
\newcommand{\bDelta}{\bar{\Delta}}

\newcommand{\be}{\begin{equation}}
\newcommand{\ee}{\end{equation}}
\newcommand{\ignore}[1]{{}}

\newtheorem{theorem}{Theorem}
\newtheorem{proposition}{Proposition}

\begin{document}
%
\title{Online Nonparametric Anomaly Detection based on Geometric Entropy Minimization}

\author{\IEEEauthorblockN{Yasin Y{\i}lmaz}
\IEEEauthorblockA{Department of Electrical Engineering\\
University of South Florida\\
Tampa, FL 33620\\
Email: yasiny@usf.edu}
}


%


\maketitle

\begin{abstract}
We consider the online and nonparametric detection of abrupt and persistent anomalies, such as a change in the regular system dynamics at a time instance due to an anomalous event (e.g., a failure, a malicious activity). Combining the simplicity of the nonparametric Geometric Entropy Minimization (GEM) method with the timely detection capability of the Cumulative Sum (CUSUM) algorithm we propose a computationally efficient online anomaly detection method that is applicable to high-dimensional datasets, and at the same time achieve a near-optimum average detection delay performance for a given false alarm constraint. 
We provide new insights to both GEM and CUSUM, including new asymptotic analysis for GEM, which enables soft decisions for outlier detection, and a novel interpretation of CUSUM in terms of the discrepancy theory, which helps us generalize it to the nonparametric GEM statistic. We numerically show, using both simulated and real datasets, that the proposed nonparametric algorithm attains a close performance to the clairvoyant parametric CUSUM test. 
\end{abstract}


%
\IEEEpeerreviewmaketitle

\section{Introduction}
\label{sec:intro}

Anomaly detection is an important problem with various applications, such as cybersecurity, quality control, fraud detection, fault detection, and health care \cite{Chan09}. It deals with identifying patterns that deviate from a nominal behavior. Although not exactly the same, it is sometimes considered equivalent to outlier detection, which deals with how well a given data sample fits to the nominal behavior \cite{Hero2006,Sri11}.
Sequential detection methods, on the other hand, considers also the temporal information in the sequence of data samples regarding a possible anomaly. In particular, sequential change detection looks for abrupt and persistent changes \cite{Basseville1993}. 
Moreover, sequential methods provide timely detection capabilities.

Parametric methods assume probabilistic models for both nominal or anomalous states. In practice, it is difficult to know the anomalous distribution, hence model mismatch limits the applicability of parametric approaches only to specific types of anomaly whose probability distribution is well aligned with the assumed model. To that end, statistical outlier detection deals only with significant deviations from the nominal distribution. However, in high-dimensional problems, it is even difficult to know the nominal distribution. Hence, recently, effective nonparametric methods for anomaly detection based on minimum entropy sets are proposed \cite{Hero2006,Sri11}. Nevertheless, these methods are solely based on outlier detection, and thus cannot use the temporal information regarding anomaly in the data samples. 

In this paper, for timely and accurately detecting abrupt and persistent anomalies, we develop a nonparametric method that is simple enough to work with high-dimensional datasets. Providing key insights to both outlier detection and change detection we merge the desirable properties of nonparametric methods and online methods. Specifically, we show asymptotic links between nonparametric Geometric Entropy Minimization (GEM) and parametric outlier detection; and also provide a novel interpretation of the sequential Cumulative Sum (CUSUM) method using the discrepancy theory, which dates back to Hermann Weyl's work in 1916 \cite{Weyl}. 

In Section \ref{sec:problem}, we formulate the problem and provide relevant background information. Then, in Section \ref{sec:odit}, we present our proposed anomaly detection method, and in Section \ref{sec:results} evaluate its performance through numerical experiments. Finally, Section \ref{sec:conc} concludes the paper.

\section{Problem Formulation and Background}
\label{sec:problem}

We monitor a system online by getting sequential observations $\mathcal{X}_t = \{X_1,X_2,\ldots,X_t\}$ of $d$-dimensional independent vectors $X_t$ in time. In this paper, we model an anomaly as persistent outliers in the observations, and aim to accurately detect such anomalies in a timely fashion using a practical nonparametric approach applicable to high-dimensional datasets (e.g., big data problems).

\subsection{Change Detection}
\label{sec:cusum}

Consider an abrupt and persistent change in the probability distribution of the observations at an unknown point in time. Let $f_0$ denote the nominal probability distribution of $X_t$ before any change, and $f$ denote the actual distribution of $X_t$.
We formulate the problem as a composite binary hypothesis test as
\begin{align}
\label{eq:ch-det}
\begin{split}
	\Hyp_0&: f=f_0, ~ \forall t \\
	\Hyp_1&: f=f_0, ~ t<\tau, ~\text{and}~ f\not=f_0, ~t \ge \tau,
\end{split}
\end{align}
where $\tau$ denotes the unknown change time. As seen in the formulation, in the change detection problem, $\Hyp_0$ (i.e., nominal distribution) is assumed true at the beginning, and the objective is to statistically detect the potential switching of the true distribution to an anomalous distribution. Hence, the change detection problem is fundamentally different than the standard binary hypothesis testing problem, in which
\begin{equation}
\label{eq:hyp-test}
	\Hyp_0: f=f_0, ~~
	\Hyp_1: f\not=f_0,
\end{equation}
and the objective is to find the true hypothesis ($\Hyp_0$ or $\Hyp_1$) from the beginning.

Standard hypothesis testing (either fixed-sample-size or sequential \cite{Poor1994}) for anomaly detection (see  \eqref{eq:hyp-test}) outputs a decision: nominal ($\phi = \Hyp_0$) or anomalous ($\phi = \Hyp_1$), whereas in change detection (see \eqref{eq:ch-det}) testing continues until the output decision is $\Hyp_1$, i.e., until an anomaly is detected \cite{Basseville1993}. Accordingly, the objective in the former is to maximize the detection probability $\Pro(\phi = \Hyp_1 | \Hyp_1)$ while satisfying a false alarm (i.e., false positive) constraint $\Pro(\phi = \Hyp_1 | \Hyp_0) \le \alpha$, whereas the objective in the latter is to minimize the \emph{average detection delay} while again satisfying a false alarm constraint. Hence, the term \emph{quickest detection} is also used \cite{Basseville1993}.

The minimax performance criterion \cite{Lorden1971}
\begin{align}
\label{eq:minimax}
\begin{split}
	\inf_T \sup_{\tau} \text{ess}\sup_{\mathcal{X}_\tau} &\Exp_{\tau}[(T-\tau)^+ | \mathcal{X}_{\tau}] \\
	\text{subject to}~ &\Exp_{\infty}[T] \ge \beta,
\end{split}
\end{align}
is commonly used to formulate the change detection problem. In \eqref{eq:minimax}, ``ess sup" denotes essential supremum, a concept in measure theory that is in practice equivalent to supremum; $(\cdot)^+ = \max\{\cdot,0\}$; $\Exp_{\tau}$ is the expectation with respect to the change time; and accordingly $\Exp_{\infty}$ is the expectation when there is no change at all.
The minimax performance criterion in \eqref{eq:minimax} minimizes the worst-case average detection delay subject to a false alarm constraint, represented by a lower bound $\beta$ on the expected alarm time when there is never a change. 

It is known that the widely used cumulative sum (CUSUM) algorithm \cite{Page1954}, is optimum with respect to \eqref{eq:minimax} when both the nominal distribution $f_0$ and the anomalous distribution $f_1$, to which $f$ switches at time $\tau$, are completely known \cite{Basseville1993}. The CUSUM procedure is given by
\begin{align}
\label{eq:cusum}
\begin{split}
	T_c &= \min\{ t: \max_{1 \le j \le t} S_j^t \ge h_c \}, \\
	S_j^t &= \sum_{i=j}^t \log \frac{f_1(X_i)}{f_0(X_i)},
\end{split}
\end{align}
where $T_c$ is the stopping time, $S_j^t$ is the running log-likelihood ratio from time $j$ to time $t$, and $h_c$ is a threshold selected to satisfy the false alarm constraint in \eqref{eq:minimax} with equality, i.e., $\Exp_{\infty}[T] = \beta$. Operationally, CUSUM stops the first time the evidence against $\Hyp_0$ is sufficiently large, achieving \emph{quickest detection} among its competitors satisfying the same false alarm constraint. However, CUSUM is a parametric method which requires the knowledge of $f_0$ and $f_1$ up to the parameters, limiting its use in high-dimensional problems where $d \gg 2$. In generalized CUSUM, which estimates the parameters of $f_0$ and $f_1$ using maximum likelihood estimation, only \emph{asymptotic optimality} is achievable \cite{Basseville1993}. To tackle high-dimensional problems we resort to nonparametric methods, such as GEM, which is discussed next.

\subsection{Outlier Detection via GEM}
\label{sec:gem}

Parametric change detection methods, in particular CUSUM, enable timely detection of certain anomaly types in which the anomalous distribution is known, as well as the nominal (i.e., baseline) distribution (e.g., change in the mean or variance of a Gaussian distribution). Outlier detection deals with the general problem of detecting unknown anomaly types (see \eqref{eq:ch-det} and \eqref{eq:hyp-test}) by considering only the likelihood under the nominal distribution. For instance, a data point is declared as an outlier if it lies outside the most compact set of data points under the nominal distribution, called the \emph{minimum volume set}. 

The minimum volume set of level $\alpha$ is given by
\begin{equation}
\label{eq:mv}
	\Omega_\alpha = \arg\min_{\mathcal{A}} \int_{\mathcal{A}} \text{d}x ~~\text{subject to}~~ \int_{\mathcal{A}} f_0(x) \text{d}x \ge 1-\alpha,
\end{equation}
where $x$ is a data point, i.e., a realization of the random variable $X_t$, $\mathcal{A}$ is the acceptance region for $\Hyp_0$ in which a data point is deemed nominal, and $\alpha$ is the significance level, i.e., constraint  on the false alarm probability. In \eqref{eq:mv}, $\Omega_\alpha$ minimizes the Lebesgue measure (i.e., volume) in $\mathbb{R}^d$ among the subsets of data points satisfying the same false alarm constraint $\alpha$ to minimize the interference with anomalous data points, i.e., to minimize $\Pro(\phi=\Hyp_0 | \Hyp_1)$ and thus to maximize the detection probability $\Pro(\phi=\Hyp_1 | \Hyp_1)$. Indeed, the detector based on the minimum volume set has a strong optimality property: it is the \emph{uniformly most powerful} test when the nominal distribution $f_0$ is Lebesgue continuous and has no flat spots over its support set, and the actual distribution $f$ is a linear mixture of $f_0$ and the uniform distribution \cite{Hero2006}. 
It is also known that the minimum volume set $\Omega_\alpha$ coincides with the \emph{minimum entropy set} which minimizes the R\'{e}nyi entropy while satisfying the same false alarm constraint \cite{Hero2006}. 

In high-dimensional datasets ($d \gg 2$), even if $f_0$ is known, it is very computationally expensive (if not impossible) to determine $\Omega_\alpha$. Hence, in the literature, there are various methods for learning minimum volume sets \cite{Scott2006}. One of them, called Geometric Entropy Minimization (GEM), is shown to be very effective with high-dimensional datasets while asymptotically achieving the performance of minimum volume set \cite{Hero2006}. Specifically, from a training set $\cX^N$ of $N$ data points distributed according to $f_0$, it first forms a k-nearest-neighbor (kNN) Euclidean graph $G=(\bcX_K^N,E)$ with $K$ vertices $\bcX_K^N \in \cX^N$ and $kK$ edges $$E=\{e_{i(l)}: i=1,\ldots,K; ~l=1,\ldots,k\},$$ where the edge length $|e_{i(l)}|$ is the Euclidean distance between the $i$th data point and its $l$th nearest neighbor in the graph, and the vertices $\bcX_K^N$ are chosen by minimizing the total weighted edge length 
\begin{equation}
	\cL_k(\cX_K^N) = \sum_{i=1}^K \sum_{l=1}^k |e_{i(l)}|^\gamma,
\end{equation}
over all possible $K$-point subsets of $\cX^N$, where $\gamma>0$ is the weight. Then, with a new data point $X_t$, it recomputes the kNN graph described above over the extended set $\cX^N \cup \{X_t\}$, resulting in the updated vertices $\bcX_K^{N+1} \in \cX^N \cup \{X_t\}$. If $X_t \in \bcX_K^{N+1}$, the new data point $X_t$ is classified as nominal; otherwise anomalous.  

In \cite{Hero2006}, using the asymptotic theory of Euclidean graphs $\bcX_K^N$ is shown to converge to the minimum volume set (and accordingly the minimum entropy set) $\Omega_\alpha$ as $$\lim_{K,N \to \infty} K/N \to 1-\alpha.$$ 
Since the original GEM algorithm, which, for each $X_t$, computes $\bcX_K^{N+1}$ over all possible $K$-point subsets of $\cX^N \cup \{X_t\}$, has exponential computational complexity, a simpler variant based on bipartite kNN graph (BP-GEM) is proposed in \cite{Sri11}. BP-GEM significantly decreases the complexity of GEM from $O(d K^2 {N \choose K})$ to $O(d N^{(8+3d)/(4+2d)})$ while maintaining the theoretical guarantees of GEM \cite{Sri11}. Specifically, BP-GEM randomly partitions the training data set $\cX^N$ into two sets $\cX^{N_1}$ and $\cX^{N_2}$, where $N_1+N_2=N$, and finds the vertices $\bcX_K^{N_1} \in \cX^{N_1}$ by minimizing
\begin{equation}
\label{eq:bp-len}
	\cL_k(\cX_K^{N_1},\cX^{N_2}) = \sum_{i=1}^K \sum_{l=k-s+1}^k |e_{i(l)}|^\gamma,
\end{equation}
over all possible $K$-point subsets of $\cX^{N_1}$, where $|e_{i(l)}|$ is the Euclidean distance from point $i$ in $\cX_K^{N_1}$ to its $l$th nearest neighbor in $\cX^{N_2}$, $1 \le s \le k$ is a fixed number introduced for convenience, and $0<\gamma<d$ is the weight. This initial graph with vertices $\bcX_K^{N_1}$ is computed only once at the beginning. Then, with the inclusion of each new data point $X_t$, we do not need to redetermine the graph vertices $\bcX_K^{N_1+1}$ over $\cX^{N_1} \cup \{X_t\}$ every time, since for each $i$ in $\bcX_K^{N_1}$ the nearest neighbors $\{i(l)\}$ are selected from the separate set $\cX^{N_2}$, and thus the total edge length $\sum_{l=k-s+1}^k |e_{i(l)}|^\gamma$ does not change, as opposed to the original GEM. Instead, we only need to compute the total edge length $\sum_{l=k-s+1}^k |e_{X_t(l)}|^\gamma$ for the new point $X_t$, and 
choose the $K$ points with smallest total edge lengths from $\bcX_K^{N_1} \cup \{X_t\}$ as the new graph vertices $\bcX_K^{N_1+1}$.

\section{The Proposed Online Discrepancy Test (ODIT)}
\label{sec:odit}

Anomaly detection solely based on outlier detection suffers from the fact that an outlier, especially one that is close to being nominal, does not necessarily correspond to an anomaly. For instance, in BP-GEM, a data point whose total edge length is slightly larger than the largest one in $\bcX_K^{N_1}$ is deemed anomalous despite the small evidence to do so, i.e., it would be decided nominal if its total edge length was a little smaller. This hard-thresholding mechanism does not provide a good link between an outlier and an anomaly. Instead, we propose to accumulate the evidence supporting anomaly in each data point (whether decided as an outlier or not), similarly to the accumulation of likelihood evidence in change detection. As a result, combining the simplicity of outlier detection with the power of sequential decision making, our proposed anomaly detector 
(i) computes the easy-to-compute outlier evidence, given by the total edge length, for each new data point; and
(ii) waits for new data by accumulating the evidence from each data point until a confident anomaly alarm can be raised instead of making a hard decision based on a single data point with little evidence. 

\subsection{Analysis of BP-GEM for Outlier Detection}

The proposed detector is motivated by the theoretical foundations explained below. 

\begin{proposition}[Test statistic of BP-GEM]
\label{pro:stat}
The BP-GEM algorithm proposed in \cite{Sri11} actually treats the hypothesis testing problem 
\begin{equation}
\label{eq:out-test}
	\Hyp_0: X_t \in \Omega_\alpha, ~~
	\Hyp_1: X_t \not\in \Omega_\alpha,
\end{equation}
with the decision function
\begin{align}
\label{eq:bp-dec}
	\phi &= \left\{
	\begin{array}{ll}
	\Hyp_0 ~~\text{if}~~ D_t \le 0 \\
	\Hyp_1 ~~\text{if}~~ D_t > 0,
	\end{array} \right. \\
\label{eq:bp-stat}
	D_t &= \sum_{l=k-s+1}^k |e_{X_t(l)}|^\gamma - \sum_{l=k-s+1}^k |e_{X_{(K)}(l)}|^\gamma, \\
\label{eq:bp-asympt1}
	& \lim_{N_1,N_2 \to \infty} D_t \overset{\text{\em monotonic}}{\sim} \log \frac{f_0(x_{\alpha})}{f_0(X_t)}, \\
\label{eq:bp-asympt2}
	\text{and}~~~ & \text{\em{sign}}\left( \lim_{N_1,N_2 \to \infty} D_t \right) = \text{\em{sign}}\left( \log \frac{f_0(x_{\alpha})}{f_0(X_t)} \right),
\end{align}
where $x_{\alpha}$ is a boundary point of $\Omega_\alpha$ (see \eqref{eq:mv}), $N_1$ and $N_2$ are the size of two partitions in the training set, and the test statistic $D_t$ is the difference between the total edge lengths of the new point $X_t$ and $X_{(K)}$, the $K$th point in $\bcX_K^{N_1}$, which has the largest total edge length in $\bcX_K^{N_1}$.
\end{proposition}

\begin{IEEEproof}
The original decision rule of BP-GEM (given as choose $\Hyp_0$ if $X_t \in \bcX_K^{N_1+1}$ and choose $\Hyp_1$ otherwise) can be restated in terms of a test statistic. 
To see this, note that $X_t \in \bcX_K^{N_1+1}$ when the new point replaces the $K$th point in the previous best set of vertices $\bcX_K^{N_1}$, i.e., its total edge length is smaller than that of the $K$th point, as shown in \eqref{eq:bp-stat}.
The asymptotic properties in \eqref{eq:bp-asympt1} and \eqref{eq:bp-asympt2} follow from the asymptotic optimality of BP-GEM. It is known \cite{Sri11} that the $\Hyp_0$-region of \eqref{eq:bp-dec} converges to the minimum volume set $\Omega_\alpha$, whose decision rule can be stated as $\Hyp_0$ if $f_0(X_t) \ge f_0(x_{\alpha})$, i.e., $\log \frac{f_0(x_{\alpha})}{f_0(X_t)} \le 0$, and $\Hyp_1$ otherwise, hence the sign property in \eqref{eq:bp-asympt2}. Assume that, as $N_2 \to \infty$, also $k \to \infty$ such that the total edge length $L_k(X_t)$ of a point $X_t$ remains a constant. In that case, $L_k(X_{t_1}) < L_k(X_{t_2})$ for all $X_{t_1}$ and $X_{t_2}$ such that $f_0(X_{t_1}) > f_0(X_{t_2})$. Since 
\be
\label{eq:stat-diff}
	D_t = L_k(X_t) - L_k(X_{(K)}), 
\ee
we have the monotonicity property stated in \eqref{eq:bp-asympt1}. Note also that $X_{(K)} \to x_\alpha$ as $N_1, K \to \infty$ such that $K/N_1 = 1-\alpha$.
\end{IEEEproof}

Proposal \ref{pro:stat} shows the structural resemblance of $D_t$ to the log-likelihood ratio between the boundary point $x_\alpha$ and $X_t$. To see the physical relationship consider the case where $f_0$ is from the exponential family, i.e., $f_0=e^{-\delta(X,\theta)}$ where $\theta$ is the parameter vector and $\delta(X,\theta)$ is a distance term causing the exponential decay in the probability density function. In this case, $\log \frac{f_0(x_{\alpha})}{f_0(X_t)} = \delta(X_t,\theta)-\delta(x_\alpha,\theta)$ is a distance metric that is similar to $D_t$ as shown by \eqref{eq:stat-diff}. They also asymptotically share a very similar structure (see Proposal \ref{pro:stat}).

This theoretical similarity between $D_t$ and the log-likelihood ratio $\log \frac{f_0(x_{\alpha})}{f_0(X_t)}$ motivates us to use the nonparametric BP-GEM approach in online anomaly detection, in a similar fashion the parametric CUSUM algorithm uses log-likelihood ratio (see \eqref{eq:cusum}).

\subsection{Online Nonparametric Anomaly Detection}

Instead of classifying each point by hard-thresholding $D_t$ as in \eqref{eq:bp-dec} and treating each outlier as an anomaly, we model an anomaly as persistent outliers in the observations and treat $D_t$ as a positive/negative evidence for anomaly. For timely and accurate detection, we propose to accumulate such anomaly evidence in time, i.e., use also the history, as opposed to the original GEM approach (see \eqref{eq:bp-dec}). Specifically, due to the independence of data points in time, we sum $D_t$ to obtain the running evidence $\Delta_t = \sum_{i=1}^t D_i$, similar to the running log-likelihood ratio in CUSUM, given by \eqref{eq:cusum}. 

The similarity between $\Delta_t$ and $S_t = \sum_{i=1}^t \log \frac{f_1(X_i)}{f_0(X_i)}$ due to Proposition \ref{pro:stat} motivates us to develop an online anomaly detector using the nonparametric test statistic $\Delta_t$. To that end, we first introduce a novel interpretation of CUSUM, which enables the generalization of the CUSUM procedure to our nonparametric test statistic.

\begin{theorem}[Discrepancy and CUSUM]
\label{thm:disc}
The CUSUM procedure, given by \eqref{eq:cusum}, can be written as
\begin{align}
\label{eq:disc}
\begin{split}
	T_c &= \min\{ t: g(\ell_t) \ge h_c \}, \\
	\ell_t &= \left[\log \frac{f_1(X_1)}{f_0(X_1)} \ldots \log \frac{f_1(X_t)}{f_0(X_t)} \right], \\
	g(\ell_t) &= \max_{1 \le n_1 \le n_2 \le t} \sum_{i=n_1}^{n_2} \ell_t^i,
\end{split}
\end{align}
where $g(\cdot)$ is a discrepancy function defined for a number sequence, similarly to the discrepancy norm defined in \cite{Moser14}, and $\ell_t^i$ is the $i$th element of the log-likelihood ratio vector $\ell_t$.
\end{theorem}

\begin{figure}[t]
\centering
\includegraphics[width=.45\textwidth]{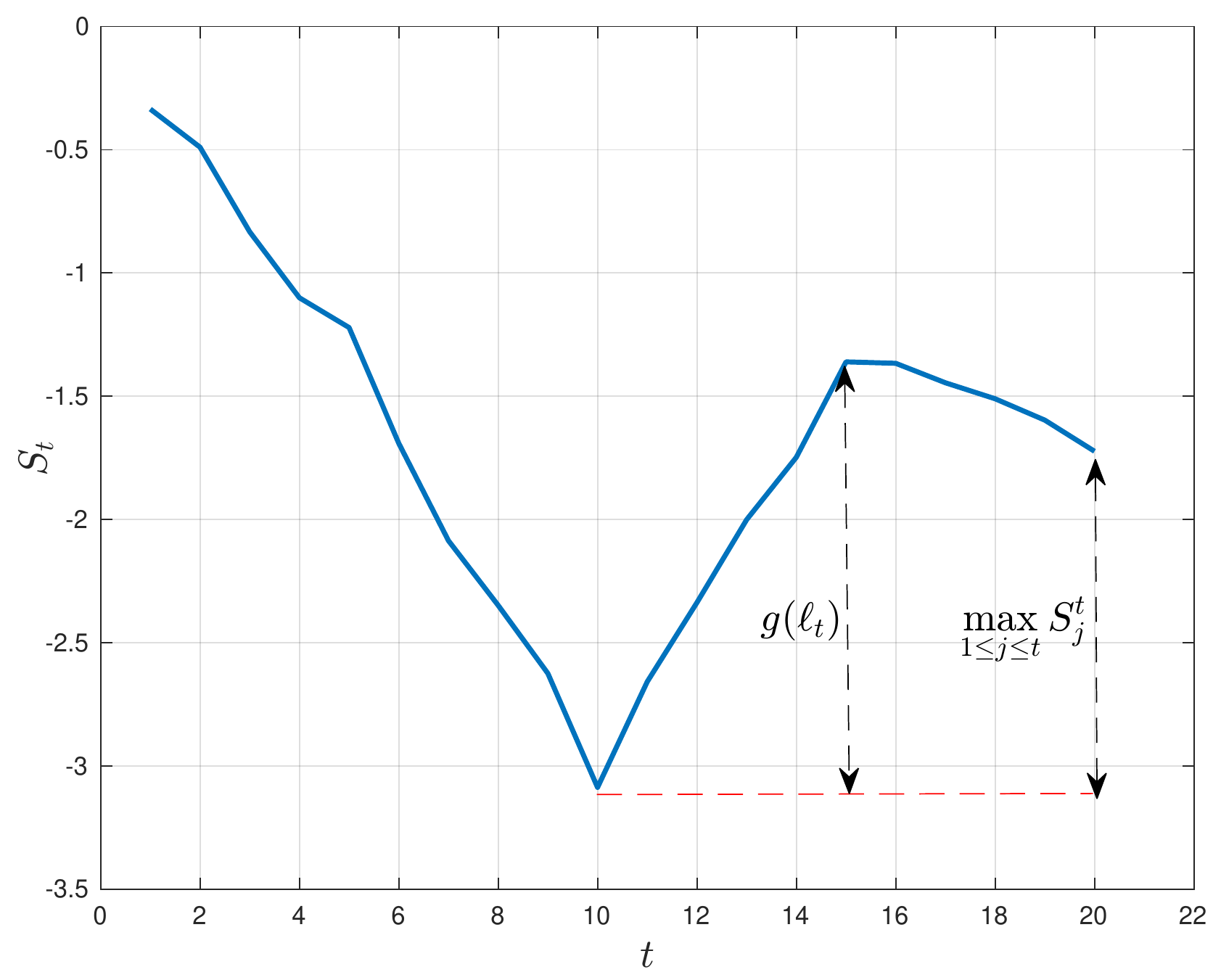}
\caption{The CUSUM statistics $\max_{1 \le j \le t} S_j^t$ and $g(\ell_t)$ from the original formulation \eqref{eq:cusum} and the alternative formulation \eqref{eq:disc}, respectively.}
\label{fig:thm}
\end{figure}

\begin{IEEEproof}
Note that in \eqref{eq:cusum} 
$$\max_{1 \le j \le t} S_j^t = \max_{1 \le j \le t} \sum_{i=j}^t \log \frac{f_1(X_i)}{f_0(X_i)} = \max_{1 \le j \le t} \sum_{i=j}^t \ell_t^i,$$
which is similar but not identical to $g(\ell_t)$ since $n_2$ in \eqref{eq:disc} is not necessarily equal to the current time $t$. Actually, $\max_{1 \le j \le t} S_j^t$ is the accumulated log-likelihood ratio since the time the minimum value of $S_t$ took place until time $t$, i.e., $S_t-S_{t_{min}}$, whereas $g(\ell_t)$ is the difference between the maximum and minimum values of $S_t$, i.e., $S_{t_{max}}-S_{t_{min}}$, as shown in Fig. \ref{fig:thm}. However, note that the original procedure, given by \eqref{eq:cusum}, stops the first time $\max_{1 \le j \le t} S_j^t$ exceeds a threshold $h_c$, which always occurs at the maximum point after the minimum occurs. That is, at the stopping time $\max_{1 \le j \le t} S_j^t = g(\ell_t)$, i.e., the stopping times obtained through \eqref{eq:cusum} and \eqref{eq:disc} are identical. 
\end{IEEEproof}

The utility of Theorem \ref{thm:disc} is that it expresses the CUSUM procedure in terms of a general discrepancy metric which is applicable to any number sequence. Specifically, according to Theorem \ref{thm:disc}, CUSUM stops when the discrepancy, measured by $g(\ell_t)$, of the observations with respect to $f_0$ is large enough. On the other hand, the original CUSUM procedure is completely based on the log-likelihood ratio, and thus not readily applicable to the ODIT statistic $\Delta_t$, which is similar, but not the same as the running log-likelihood ratio $S_t$.

Our discrepancy function $g(\cdot)$ is motivated by the discrepancy theory \cite{Weyl}, and defined similarly to the discrepancy norm in \cite{Moser14}. The function in \cite{Moser14} measures the maximum bidirectional change (i.e., increase or decrease) in a vector, whereas our function measures the maximum increase in a sequence/vector. 

Before presenting the proposed ODIT algorithm, we recall the CUSUM formulation that recursively updates its test statistic \cite{Basseville1993},
\begin{align}
\label{eq:cusum_recursive}
\begin{split}
	T_c &= \min\{ t: \bar{S}_t \ge h_c \}, \\
	\bar{S}_t &= \max\left\{\bar{S}_{t-1}+\log \frac{f_1(X_t)}{f_0(X_t)}, 0 \right\}, ~~ \bar{S}_0 = 0.
\end{split}
\end{align}
In computing the discrepancy function, since we are only interested in the increase from the minimum value in the past, as shown in Fig. \ref{fig:thm}, in the proposed algorithm we clip the minimum value always at zero as in \eqref{eq:cusum_recursive}, which yields the following stopping rule and the easy-to-compute recursive update rule for the ODIT statistic
\begin{align}
\label{eq:odit}
\begin{split}
	T_d &= \min\{ t: \bDelta_t \ge h \}, \\
	\bDelta_t &= \max\{\bDelta_{t-1}+D_t,0\}, ~~ \bDelta_0 = 0.
\end{split}
\end{align}

The proposed ODIT anomaly detector is summarized in Algorithm \ref{alg}.

\begin{algorithm}                      
\caption{The proposed anomaly detector: ODIT}          
\label{alg}                           
\begin{algorithmic}[1]                    
    \STATE {\bf Initialize:} $\bDelta=0$, $t \gets 1$
    \STATE Partition training set into $\cX^{N_1}$ and $\cX^{N_2}$
    \STATE Determine $\bcX_K^{N_1}$ as in \eqref{eq:bp-len}  
        \WHILE{$\bDelta < h$}
        \STATE Get new data $X_t$ and compute $D_t$ as in \eqref{eq:bp-stat}
        \STATE $\bDelta = \max\{\bDelta+D_t,0\}$
        \STATE $t \gets t+1$ 
    \ENDWHILE
   \STATE Declare anomaly
\end{algorithmic}
\end{algorithm}

\ignore{
The update equation in line 6 enables the recursive computation of the test statistic $\|\Delta_t\|_{\mathcal{D}}$ where $\Delta_t = [D_1,\ldots,D_t]^T$ is the vector of $D_t$, given in \eqref{eq:bp-stat}.
}

\section{Numerical Results}
\label{sec:results}

\begin{figure}[t]
\centering
\includegraphics[width=.4\textwidth]{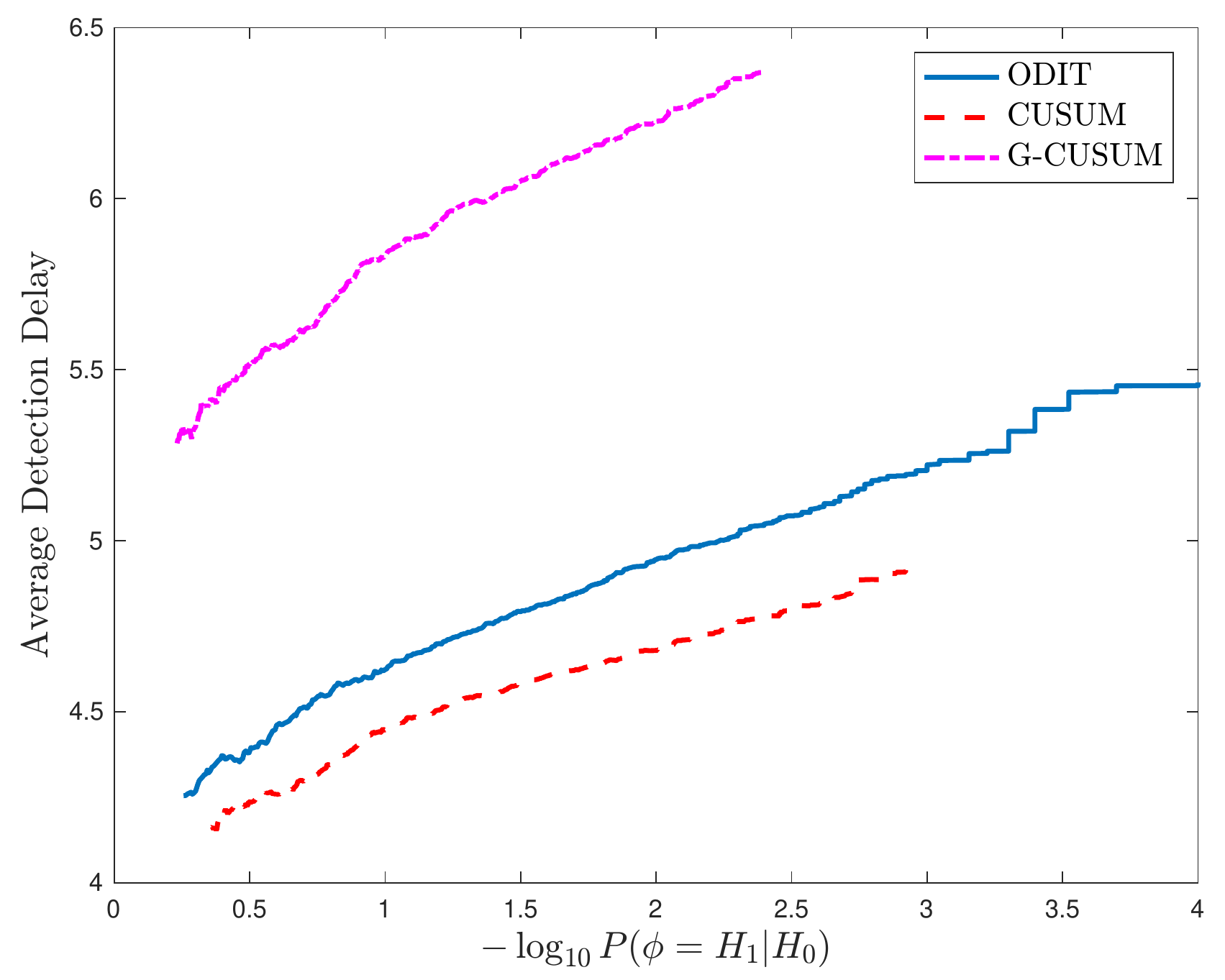}
\caption{Simulated data with a Gaussian nominal distribution $f_0$ and an anomalous distribution $f_1$ which is a mixture of nominal and uniform distributions. Average detection delay vs. false alarm probability performances of the proposed nonparametric ODIT detector, the parametric clairvoyant CUSUM detector, which knows both $f_0$ and $f_1$ exactly, and the generalized CUSUM detector, which exactly knows $f_0$, but estimates the uniform distribution upper bound as $0.9$ instead of the true value $1$ in $f_1$.}
\label{fig:add}
\end{figure}

In this section, we provide numerical results to compare the proposed nonparametric ODIT detector with the parametric clairvoyant CUSUM  detector and the generalized CUSUM detector, which estimates the model parameters. Following the simulations in \cite{Hero2006} and \cite{Sri11}, we first test the scenario in which the nominal distribution $f_0$ is a 2-dimensional Gaussian with zero mean and diagonal covariance with standard deviation $\sigma=0.1$; and the anomalous distribution $f_1$ is a mixture of $f_0$ and 2-dimensional uniform distribution on $[0,1]^2$: $f = 0.8 f_0+0.2 U$. The training set consists of $N=10000$ points with partitions $N_1=1000$ and $N_2=9000$. The test set contains $500$ points with anomaly time at $100$. Thus, after $t=100$, we see an anomalous point on average every five time instance. We set $\alpha=0.05$, $K=\alpha N_1$, $k=1$, and $s=1$. 
As seen in Fig. \ref{fig:add}, the proposed nonparametric detector well approximates the parametric optimum CUSUM detector, which knows $f_0$ and $f_1$ exactly.
It achieves near-optimum performance while being computationally simple and free of assumptions on the nominal and anomalous probability distributions. Furthermore, the proposed detector significantly outperforms the generalized CUSUM (G-CUSUM), which is used in practice as it estimates the unknown parameters using the maximum likelihood approach. Note the significant performance gap even with a small error in estimating the anomalous distribution. 

We also test our proposed algorithm on the ``Heterogeneity Human Activity Recognition Dataset" \cite{Stisen15} obtained from the UCI Machine Learning Repository \cite{UCI}. This dataset contains accelerometer and gyroscope data from smartphones and smartwatches about 6 activities: biking, sitting, standing, walking, stair up, and stair down. We used the smartwatch accelerometer data, which contains around 3.5 million data points with 5 numeric features. Focusing on the activity transitions we tested the online detection performance of our algorithm in terms of average detection delay vs. false alarm probability, and compared it to that of G-CUSUM which fits multivariate Gaussian models to $f_0$ and $f_1$ by estimating the parameters from some training data using the maximum likelihood approach. 

\begin{figure}[t]
\label{fig:data}
\centering
\includegraphics[width=.4\textwidth]{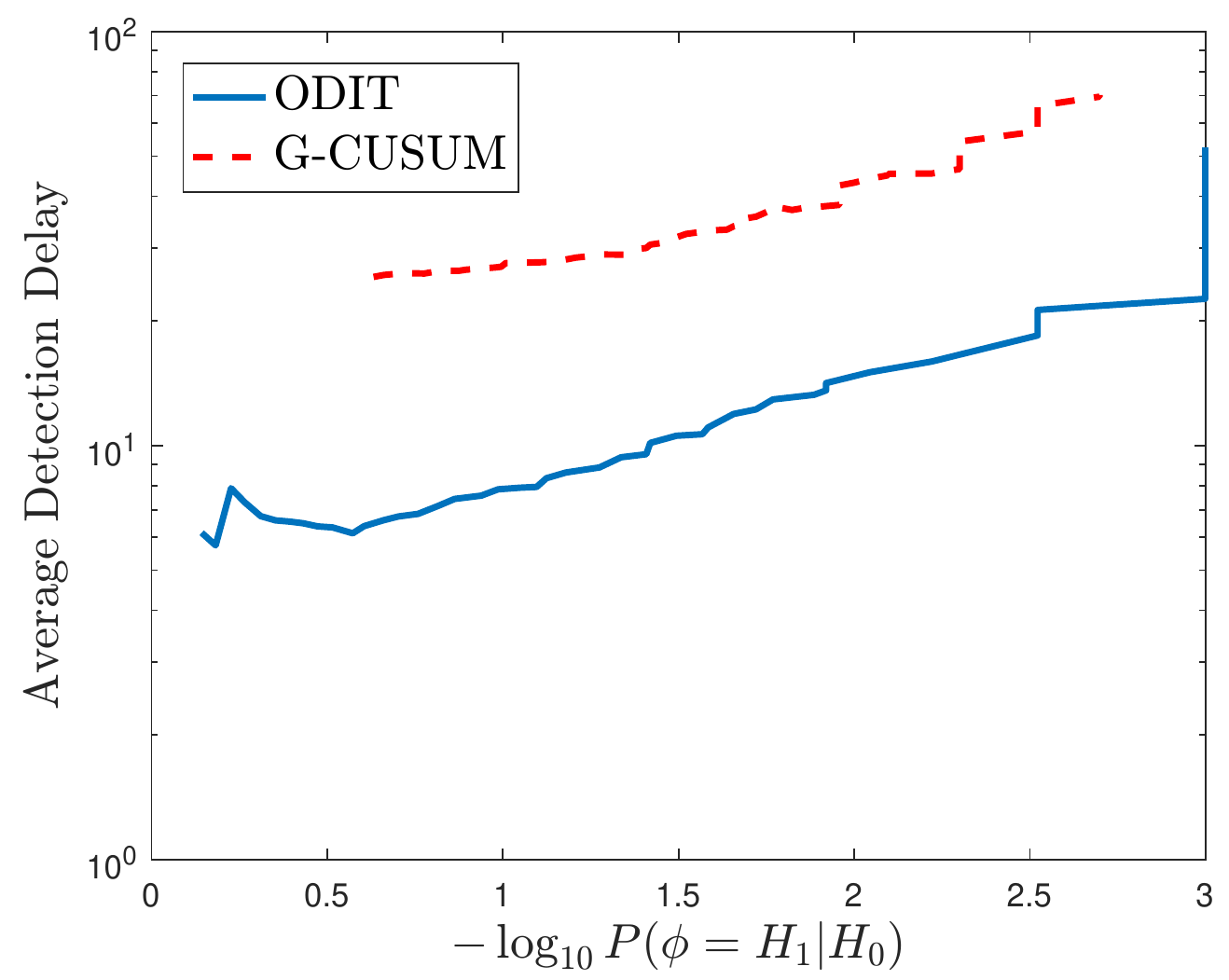}
\caption{Performance comparison with the Heterogeneity Human Activity Recognition Dataset. Average detection delay vs. false alarm probability performances of the proposed nonparametric ODIT detector and the parametric generalized CUSUM detector.}
\end{figure}

\section{Conclusion}
\label{sec:conc}

We developed a computationally efficient nonparametric method for timely detection of abrupt and persistent anomalies. The proposed algorithm combines the GEM outlier detection approach with the CUSUM sequential change method. New insights into GEM and CUSUM were provided to effectively combine them. The introduced relation between CUSUM and the discrepancy theory enables extensions of the CUSUM procedure to new applications, as we showed for nonparametric anomaly detection. Similarly, the relation of GEM to parametric outlier detection enabled us to use the simple nonparametric outlier detection approach for timely detection of statistical changes. Numerical results on both simulated and real datasets justified the effectiveness of the proposed algorithm.


%




\ifCLASSOPTIONcaptionsoff
  \newpage
\fi

\end{document}